\documentclass[useAMS,usenatbib]{mn2e}
\usepackage{graphicx, natbib, amssymb, amsmath}
\title[Detecting highly-dispersed bursts with next-generation radio telescopes]{Detecting highly-dispersed bursts with next-generation radio telescopes}
\author[Hassall, Keane and Fender]{T.~E.~Hassall$^1$\thanks{Email: t.hassall@soton.ac.uk}, E.~F.~Keane$^{2,3}$ and R.~P.~Fender$^1$\\
$^1$ School of Physics and Astronomy, University of Southampton, Southampton, SO17 1BJ, UK\\
$^2$ Jodrell Bank Centre for Astrophysics, School of Physics and Astronomy, The University of Manchester, Manchester M13 9PL, UK\\
$^3$ Centre for Astrophysics and Supercomputing, Swinburne University of Technology, P.O. Box 218, Hawthorn, VIC 3122, Australia}
\begin{document}

\date{}

\pagerange{\pageref{firstpage}--\pageref{lastpage}} \pubyear{2013}

\maketitle

\label{firstpage}

\begin{abstract}
Recently, there have been reports of six bright, dispersed bursts of coherent radio emission found in pulsar surveys with the Parkes Multi-beam Receiver. Not much is known about the progenitors of these bursts, but they are highly-energetic, and probably of extragalactic origin. Their properties suggest extreme environments and interesting physics, but in order to understand and study these events, more examples need to be found. Fortunately, the recent boom in radio astronomy means many `next-generation' radio telescopes are set to begin observing in the near future. In this paper we discuss the prospects of detecting short extragalactic bursts, in both beamformed and imaging data, using these instruments. We find that often the volume of space probed by radio surveys of fast transients is limited by the dispersion measure (DM) of the source, rather than its physical distance (although the two quantities are related). This effect is larger for low-frequency telescopes, where propagation effects are more prominent, but, their larger fields-of-view are often enough to compensate for this. Our simulations suggest that the low-frequency component of SKA$_1$ could find an extragalactic burst every hour. We also show that if the sensitivity of the telescope is above a certain threshold, imaging surveys may prove more fruitful than beamformed surveys in finding these sorts of transients.
\end{abstract}

\begin{keywords}
scattering -- methods: observational -- surveys -- intergalactic medium -- galaxies: ISM 
\end{keywords}

\section{Introduction}
Fast radio bursts (FRBs) are amongst the most violent and energetic events in the Universe. To date, there have been six bright, highly-dispersed bursts found in beamformed surveys for fast transients \citep{lbm+07, kkl+11, tsb+13}. The progenitors of the bursts are unknown, but the limited data that we have on them point towards extreme environments and interesting physics. Their dispersion measures (DMs), indicate that they are likely to be from outside of our Galaxy, and probably at cosmological distances (their inferred redshifts are $z\sim0.1-0.9$). Although their exact distances ($D$) and intrinsic pulse widths ($\tau_0$) are not well constrained, all of the bursts are known to be highly energetic, coherent emitters -- FRB~010724 had a brightness temperature $\sim 10^{34}\left(\frac{D}{500~\mathrm{Mpc}} \right)^2\left(\frac{\tau_0}{5~\mathrm{ms}}\right)^{-2}$~K and released $\sim10^{33}\left(\frac{D}{500~\mathrm{Mpc}} \right)^2\left(\frac{\tau_0}{5~\mathrm{ms}}\right)$~J of energy. 
Light-travel-time arguments show that the bursts must originate from compact regions (with upper limits on the diameters ranging from $300 - 1500$~km). \cite{ksk+12} argued that FRB~010621 could potentially be explained as an annihilating mini black hole or a `giant pulse' from a young pulsar with a low burst-rate, but only if the NE2001 model of the Galactic electron-density \citep{cl02} is sufficiently incorrect as to allow the burst to originate from inside our Galaxy. They found no consistent explanation for FRB~010724, although \cite{lbm+07} noted that the implied rate of occurrence is compatible with that of gamma-ray bursts. 

As well as being interesting in their own right, these objects are also potentially powerful tools for studying the intergalactic medium (IGM). They vary on very short timescales, allowing us to measure the dispersive delay and scatter broadening timescale, which can be used to determine the density and spatial distribution of electrons along the line-of-sight. They may also prove useful for cosmological measurements if the bursts are standard, or `standardizable', \citep[][]{phi93} candles.

In order to study and understand this population of objects, it will be necessary to find many more of them. All of the known bursts have been found through `pulsar-like' beamformed observations at relatively high frequencies ($\sim$1.4~GHz), but, because FRBs are so bright, they may also be detectable in images. Note, in this paper we choose to use 'beamformed' to describe pulsar-like observations, even though in some cases beams are not formed digitally. Beamformed surveys have been successful in the past, and they are known to be sensitive to these bursts and to other bright single pulses \citep[e.g. Rotating Radio Transients, `RRATs',][]{km11}. However, the localisation of the sources is relatively poor, and in some cases, it can be difficult to distinguish real pulses from radio-frequency interference \citep[RFI, see][]{bbe+11}. Imaging surveys could (in some circumstances) offer a better localisation of the source, and are also potentially more robust against RFI, but because the bursts are so rapid and radio images have a practical limit on their shortest exposures, some of the sensitivity of the survey may be lost. In this paper, we compare the number of simulated FRBs detected in imaging and beamformed surveys with several `next-generation' radio telescopes. We also investigate the effects of dispersion and scattering on the bursts and determine the most effective observatories for locating such events.

\section{Propagation Effects}
As pulsed emission passes through an ionised plasma it interacts with electrons along the line-of-sight, and becomes distorted by dispersion and scattering. The peak flux of a pulse is reduced, and the emission becomes smeared out in time, making it more difficult to detect. Although propagation effects are particularly prominent at low frequencies, they need to be considered at all frequencies when trying to determine the rate of short bursts. In the following section we summarise the two propagation effects we consider here.

\subsection{Dispersion}
The frequency-dependent refractive index of a cold, ionised plasma, means that any signals propagating through it are dispersed. Emission at frequency, $\nu$ (in MHz), is delayed with respect to emission at infinite frequency by, $\Delta t_\mathrm{DM}$ (in s):
\begin{equation}
\Delta t_\mathrm{DM} = \frac{\mathrm{DM}}{2.410\times10^{-4}\nu^2} ,
\label{eq:disp}
\end{equation}
where DM is the dispersion measure in units of pc~cm$^{-3}$. In beamformed analysis, this delay can be addressed either by channelising the data and compensating for the delay in each channel, significantly reducing the effects of dispersion across the band \citep[`incoherent dedispersion',][]{lv71}, or applying a frequency-dependent delay to the raw voltage data directly and completely removing the dispersive delay \citep[`coherent dedispersion',][]{hr75}. Coherent dedispersion is the more precise method, but typically incoherent dedispersion is used in blind searches for pulsars and fast transients, as coherent dedispersion is usually too computationally expensive. 

Whilst dedispersion makes searching for new objects more difficult, once the correct DM of a source is known, it can be removed. Dispersion also provides a good way to discriminate between a real signal and radio frequency interference (RFI), as terrestrial signals do not typically follow a $\nu^{-2}$ law. The fact that all of the previously reported bursts follow the dispersion law so well, remains the strongest evidence that they are of astrophysical origin. As the dispersive delay depends only on the number of electrons along the line-of-sight, and not their distribution, the dispersive delay will follow the same $\nu^{-2}$ law in both the interstellar medium (ISM) and the intergalactic medium (IGM). However, the relation between the DM and the column density of electrons of redshifted sources will be different from that predicted theoretically and observed for Galactic sources \citep[see][for a more detailed discussion]{iok03}.

\subsection{Scattering}
\label{sec:scattering}

\begin{figure}
\begin{center}
\includegraphics[height=\linewidth, angle=90, trim=0.5cm 2.5cm 1.25cm 2.25cm, clip=True]{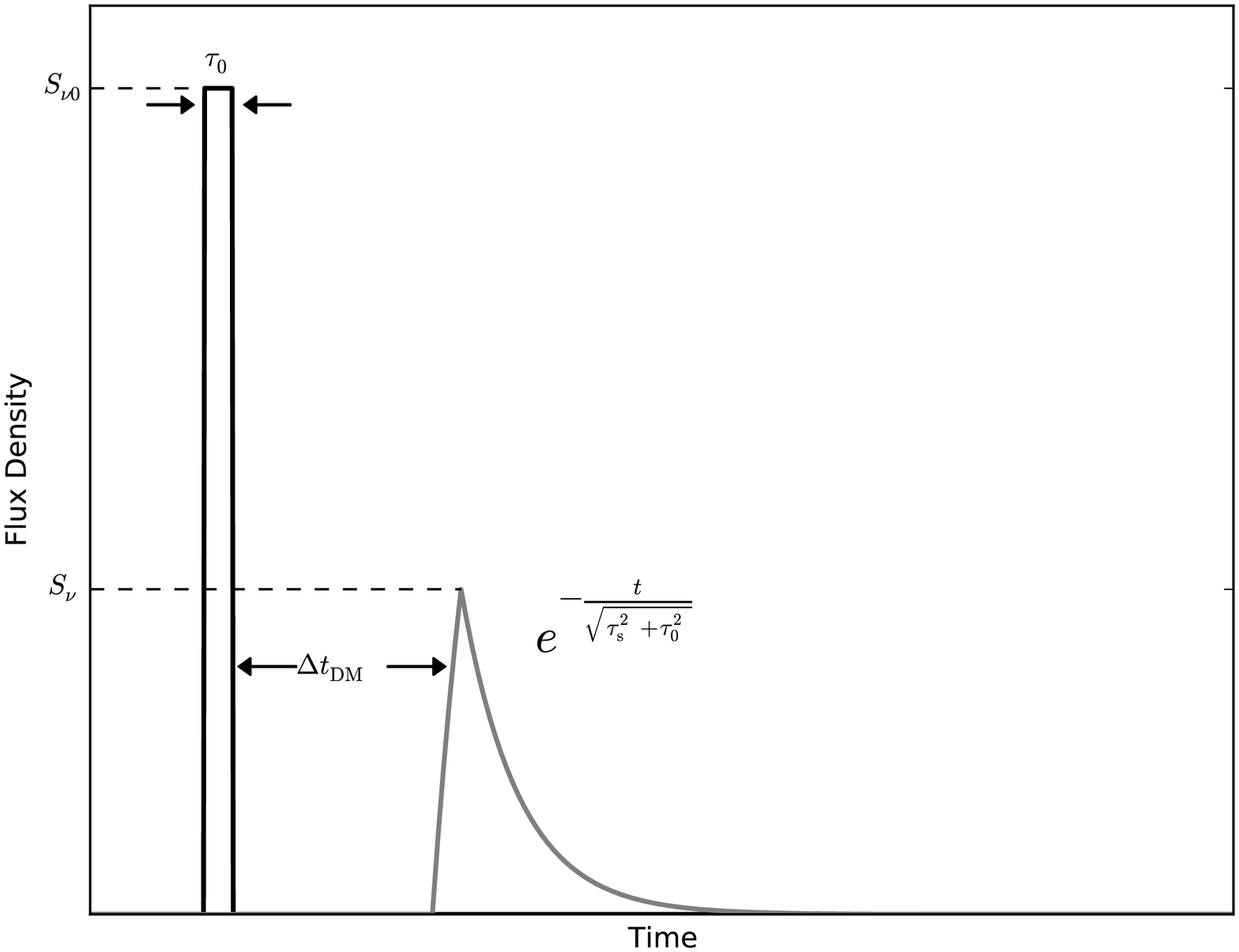}
\caption{A boxcar pulse with initial pulse density $S_{\nu 0}$ (black line) is scatter broadened by the interstellar medium. Its peak flux density is reduced to $S_\nu$ and its profile changes (grey line), increasing the pulse width to $\sim\sqrt{\tau_0^2+\tau_\mathrm{s}^2}$.}
\label{fig:fluence}
\end{center}
\end{figure}

As radio waves propagate through a plasma, they are also scattered by fluctuations in the electron density along the line-of-sight. This results in multi-path propagation, which temporally broadens narrow pulses. In the simplest case of a scattering screen located midway between the source and observer this effectively convolves the narrow pulse with a one-sided exponential with scattering timescale, $\tau_\mathrm{s}$\footnote{Multi-path propagation also blurs the pulses spatially, increasing the observed angular size of the source by $\theta_s \sim \sqrt{\frac{\tau_\mathrm{s} c}{D}}$. We do not consider this effect here, however, because the distance to the source is typically large. To blur a source 1~kpc away by 10 arcseconds, $\tau_s \sim 250$ seconds.} (see Figure~\ref{fig:fluence}). Detailed reviews of interstellar scattering can be found in \cite{ric90} and \cite{nar92}. 

The scatter-broadening time of a pulsed signal passing through the ISM is related to the DM by the empirical function derived by \cite{bcc+04}, 
\begin{equation}
\log \tau_\mathrm{s} = 2.12 + 0.154\log(\mathrm{DM}) + 1.07(\log\mathrm{DM})^2 - 3.86\log \nu ,
\label{eq:scatt}
\end{equation}
where $\nu$ is the observing frequency in MHz, and $\tau_\mathrm{s}$ is the scatter broadening time in s. Although, it should be noted that observed values of $\tau_\mathrm{s}$ for a given DM  can deviate from the predicted value by up to two orders of magnitude and the frequency-dependence of $\tau_\mathrm{s}$ depends strongly on the distribution of the electron density along the line-of-sight. 

As the peak flux is typically used as the detection threshold it is important we address how it is affected by scattering. When a pulse is scattered, it is smeared out over a longer time, reducing the peak flux density. If we assume that the unscattered pulse is a step function with intrinsic width $\tau_0$, and a peak flux density of $S_{\nu 0}$, then its intrinsic fluence is $\mathcal{F} = S_{\nu 0} \tau_0$. As the pulse is broadened by scattering, the fluence at a given frequency, $F_\nu$ is given by:
\begin{equation}
\label{eq:fluence}
 \mathcal{F}_\nu = S_\nu \sqrt{ \tau_0^2 + \left( \int_0^\infty e^{-t/\tau_\mathrm{s}} dt \right)^2 } = S_\nu \sqrt{ \tau_0^2+\tau_\mathrm{s}^2 } , 
\end{equation}
where $S_\nu$, is the observed peak flux density at frequency, $\nu$. If we assume that the fluence is conserved by scattering (i.e. any emission scattered out of the line-of-sight is balanced by emission scattered into it), then the peak flux is given by:
\begin{equation}
S_\nu = \frac{\mathcal{F}}{ \sqrt{\tau_0^2+\tau_\mathrm{s}^2} }.
\end{equation}
At high frequencies (or low DMs), where $\tau_0 >> \tau_\mathrm{s}$, the peak flux is approximately constant, but when $\tau_\mathrm{s} \gtrsim \tau_0$, scattering causes the peak flux to decrease as $S_\nu\propto1/\tau_\mathrm{s}$ ($\propto \nu^{3.86}$, if the scatter-broadening follows Equation~\ref{eq:scatt}). At low observing frequencies (or high DMs), the temporal pulse broadening is large, and there is a significant reduction in peak flux.

Scatter-broadening of extragalactic sources has only been observed twice \citep{tsb+13}, and is therefore not well understood, but current observations suggest that it is probably quite different from the scattering seen from the ISM. It has not yet been possible to measure the distribution of electrons in the IGM, but one possibility is that the density fluctuations are less significant than those seen in the ISM.  The so-called `lever-arm' effect is also expected to reduce scatter-broadening; the observed effect is maximised when scattering material is concentrated halfway between the observer and the source. So, because most of the scattering material will be concentrated in the ISM of our Galaxy and the host galaxy, the scatter-broadening of extragalactic sources may be reduced \citep{lkm+13}. This is consistent with the findings of the two known FRBs, which, despite having large DMs, showed very little scatter broadening \citep[in many cases, significantly less than what was expected from][see Table~\ref{tab:bursts}]{bcc+04}.

\subsection{Propagation Effects in Known FRBs}

\begin{table*}
\begin{minipage}{17.5cm}
\caption{Observed properties of known extragalactic bursts.}
\begin{tabular}{l|c|c|c|c|c|c}
\hline
  & FRB 010724 & FRB 010621 & FRB 110220 & FRB 110627 & FRB 110703 & FRB 120127  \\
\hline
Observed Width (ms) & 4.6& 8.3 & 5.6 & $<1.4$ & $<4.3$ & $<1.1$\\
$(\tau_{\mathrm{0}}^2 + \tau_{\mathrm{s}}^2)^{1/2}$~(ms) & $3.1$ & $4.8$   &$5.5$&$<1.1$&$<4.1$&$<0.9$  \\
Predicted $\tau_\mathrm{s}$ (ms)$^\star$ & 2.89 &  177 & 802 & 145 & 2251 & 28 \\
Dispersion Measure(pc~cm$^{-3}$) & 375$\pm1$& 746$\pm1$ & 944.38$\pm0.05$ & 723.0$\pm0.3$ & 1103.6$\pm0.7$ & 553.3$\pm0.3$\\
Extragalactic DM (pc~cm$^{-3}$) & 330 & 213 & 910 & 677 & 1072 & 521\\
Peak Flux Density (Jy) & 30$\pm10$& 0.4$\pm0.1$ & 1.3 & 0.4 & 0.5 & 0.5\\
Spectral Index$^{\dagger \ref{fn:spindx}}$ & $-4\pm1$& 0$\pm1$ & 0$\pm1$& 0$\pm1$& 0$\pm1$& 0$\pm1$\\
Observed Rate (hr$^{-1}$~deg$^{-2}$)$^\ddagger$ & $0.0019^{+0.0045}_{-0.0006}$& $0.00051^{+0.0013}_{-0.0001}$ & $0.0017^{+0.0013}_{-0.0005}$& $0.0017^{+0.0013}_{-0.0005}$& $0.0017^{+0.0013}_{-0.0005}$ & $0.0017^{+0.0013}_{-0.0005}$\\
\end{tabular}

\medskip
$^\star$From \cite{bcc+04}. \newline
$^\dagger$Spectral index of the peak flux density.\newline
$^\ddagger$Uncertainties are determined following \cite{geh86}.
\label{tab:bursts}
\end{minipage}
\end{table*}%

We find it appropriate here to compare the properties of the previously reported bursts. The relation between the observed width, $W$, and the intrinsic width, $\tau_{\mathrm{0}}$ can be written as:
\begin{equation}
  W^2 = \tau_{\mathrm{0}}^2 + \tau_{\mathrm{DM}}^2 + \tau_{\mathrm{s}}^2 + \tau_{\mathrm{samp}}^2 \; ,
\end{equation}
where $\tau_{\mathrm{DM}}$ is the dispersive delay across a single channel; 
for a frequency resolution $\Delta\nu$, an observing frequency of $\nu$ and a dispersion measure DM, the dispersive smearing is: $\tau_{\mathrm{DM}} = 4.5 (\mathrm{DM}/500\;\mathrm{cm}^{-3}\;\mathrm{pc}) (\Delta \nu/3\;\mathrm{MHz}) (1400\;\mathrm{MHz}/\nu)^3\;$ms;  $\tau_{\mathrm{s}}$ is the unknown scattering time; and $\tau_{\mathrm{samp}}$ is the sampling time of the observation. Unless the pulse is completely resolved, and the scatter-broadening can be measured \citep[as for FRB~110220, see][]{tsb+13}, the extent of the scatter-broadening is unknown and we can only place an upper limit on the intrinsic pulse width of any detected signal.

FRB~010724 (010621) had an observed pulse width of $4.6$~ms ($8.3$~ms) at an observational frequency of $1400$~MHz. Removing the dispersive smearing gives $(\tau_{\mathrm{0}}^2 + \tau_{\mathrm{s}}^2)^{1/2}$, i.e. an upper limit on the intrinsic width (at 1400~MHz), of $3.1$~ms ($4.8$~ms), with a corresponding upper limit on the size of the source of $\sim900$~km ($\sim1400$~km) which is much smaller than the minimum allowed radius of a white dwarf. Similarly, the FRBs found by \cite{tsb+13} also seem to have narrow intrinsic pulse widths, the only pulses with $W>2$~ms seem to have been scatter broadened.
In what follows we consider intrinsic pulse widths of $1$~ms. All FRBs have DM values in excess of the maximum expected contribution from the Galaxy along their respective lines-of-sight. This places them outside the Galaxy, and using the model of \citet{iok03} we can infer redshifts between $z\sim 0.1-0.9$ for the sources. Using a cosmological model (e.g. $\mathrm{\Lambda}$CDM) we can infer a distance and thence a luminosity. This line of reasoning leads to the conclusion that the bursts are very bright, $\sim12$ orders of magnitude more luminous than the typical pulses seen from pulsars, and $\sim6$ orders of magnitude more luminous than the brightest pulse ever observed from the Crab pulsar \citep[see Figure~1 of][]{kkl+11}. We note that the dependence on frequency of the flux density and the pulse width is seen to be quite steep for FRB~010724 but very flat for FRB~010621 (although if either of the bursts was detected away from the centre of the telescope beam, the spectral index would appear steeper than it actually is). The spectral indices of the bursts from \cite{tsb+13} were all consistent with being flat. In our simulations below we consider a wide range of spectral indices\footnote{In this paper, we follow the convention of defining the spectral index $\alpha$ as $S_\nu \propto \nu^{\alpha}$. \label{fn:spindx}} and show the results for all cases. Some properties of the bursts are summarised in Table~\ref{tab:bursts}.

\section{Rate Calculations}
\subsection{Determining Rates of FRBs}
As there are only six known FRBs, it has so far been impossible to properly determine the luminosity function of the bursts. So, for the purposes of this paper, we will assume that the bursts are standard candles, which emit over all frequencies following a constant spectral index. It then follows that the number of events which will be seen in a given observation ($N_\mathrm{obs}$) is given by:
\begin{equation}
N_\mathrm{obs} = \rho_0 t_\mathrm{obs}V_\mathrm{obs}~,
\label{eq:rate}
\end{equation}
where $\rho_0$ is the rate at which the events occur per unit time per unit volume, $t_\mathrm{obs}$ is the total amount of observing time, and $V_\mathrm{obs}$ is the volume of extragalactic space being probed in the observations. We note that, as FRBs are at cosmological distances,  the co-moving volume must be used for this calculation. For a given instrument, which is sensitive to bursts above a given luminosity out to a co-moving radial distance of $D_\mathrm{max}$ and has a beamshape $B$, this is given by:
\begin{equation}
V_\mathrm{obs} = \int_{\phi=0}^{2\pi} \int_{\theta=0}^\pi \int_{r=0}^{D_\mathrm{max}} B r^2 \sin \theta dr d\theta d\phi
\end{equation}
Generally the beamshape will not depend on $r$, so we can rewrite the equation as:
\begin{equation}
V_\mathrm{obs} = \frac{1}{3}D_\mathrm{max}^3 \int_{\phi=0}^{2\pi} \int_{\theta=0}^\pi  B(\theta, \phi) \sin \theta d\theta d\phi =  \frac{\Omega}{3}D_\mathrm{max}^3 ~,
\label{eq:v}
\end{equation}
where $\Omega$ is the integrated surface area of the beamshape in steradians\footnote{$\Omega \approx a^2$ if we assume the beam is a square step function, and $\Omega \approx 2\pi a^2$ for a circularly symmetrical gaussian beam with (small) angular width $a$. More complex beam patterns (for example, that of LOFAR), are most easily determined by integrating the beamshape numerically.}. Typically, in the local Universe, $D_\mathrm{max}$ varies as $S_\mathrm{min}^{-1/2}$ (the minimum-detectable flux density) which, combined with Equations~\ref{eq:rate} and \ref{eq:v}, gives rise to the well-known $N \propto S^{-3/2}$ relation. 

In radio surveys for `fast' transients, however, propagation effects are important. Dispersion delays the pulse with decreasing frequency and scattering broadens the pulse, reducing its peak flux. This means that $D_\mathrm{max}$ may be limited by the dispersion and scatter broadening along the line-of-sight rather than the luminosity and distance to the source. A significant contribution to the DM of extragalactic bursts comes from our Galaxy\footnote{The maximum Galactic DM ranges from $\sim$20--2000~pc~cm$^{-3}$ depending on the line-of-sight \citep{cl02}.} (and any putative host galaxy). Because it reduces the peak flux of the signal, this component of the DM effectively reduces the observable volume of the Universe. The rest of the DM comes from the IGM. This intergalactic DM ($\mathrm{DM}_\mathrm{IGM}$, in pc~cm$^{-3}$), may be related to the redshift of the source \citep{iok03}:
\begin{equation}
z \approx \frac{\mathrm{DM}_\mathrm{IGM}}{1000}~.
\end{equation}
Note, the value DM$_\mathrm{IGM}$ given by Ioka accounts for the fact that the observed emission has been cosmologically redshifted. This redshift can be converted into a co-moving radial distance using \textsc{CosmoCalc} \citep{wri06}, with the cosmological constants derived from the latest  \emph{Planck} results \citep[$H_0= 68~\mathrm{km~s}^{-1}, \Omega_\mathrm{M}=0.32, \Omega_\Lambda =0.68$,][]{aaa+13}. 

The dispersion and scattering in the ISM and the IGM mean that, in radio surveys, the flux density of dispersed transients will appear to drop more quickly with distance than expected from the inverse-square law. There is an additional effect which needs to be considered which arises because the optimal signal-to-noise ratio (SNR) of a given dataset is achieved when the binning time of the data is approximately equal to the pulse width. If this time is too short, then not all of the pulse is observed, and if it is too long then unnecessary noise is added to the signal, effectively increasing the minimum flux we are sensitive to (and therefore reducing the observed volume). The combination of the luminosity, distance, propagation effects and binning time makes $D_\mathrm{max}$ difficult to determine analytically, so we use simulations to determine the rate of FRBs, and the volume of the Universe sampled by several next-generation observatories.

\subsection{Simulations}
We simulated pulses (with 1~ms intrinsic width, assumed to be standard candles with the mean specific luminosity of the six known bursts; 50~Jy~Gpc$^2$ at 1400~MHz\footnote{This specific luminosity, if persisting over 20 GHz, corresponds to a radio-band luminosity of $\sim10^{37}$~W and an energy release of $10^{34} (W/1~\mathrm{ms})$~J.}) propagating through space, and interacting with the ISM and IGM, then being observed with a range of integration/binning times. For our calculations, we assume that the fluence of the pulse is conserved through scattering; if this is not the case, the sensitivity to scattered FRBs will be reduced. Binning times for beamformed observations were limited to $0.001~\mathrm{s} < t_\mathrm{bin} < 0.1~\mathrm{s}$, and the integration times for images were chosen to be $1~\mathrm{s} < t_\mathrm{int} < 10000~\mathrm{s}$.  The lower limit of 1~s for imaging surveys was arbitrarily chosen to distinguish between imaging and beamformed observations. If images are taken on shorter timescales, then the imaging rates will improve because they will be more sensitive to unscattered FRBs. The dwell times for the simulated observations were chosen to be 600~s for beamformed observations, and 10000~s for imaging observations. Care should be taken when selecting a dwell time for FRB observations to ensure the dwell time is significantly longer than the dispersive delay across the band, so the probability of detecting the whole dispersed pulse (and hence the sensitivity of the observation) is maximised. We assume that beamformed data are perfectly dedispersed, and images are not dedispersed at all. The simulations are performed for DMs from $0-6000$~pc~cm$^{-3}$ ($z\sim0-6$, where $z\sim6$ corresponds to the end of cosmic reionization in Ioka's simulations). For beamformed observations, we assume that the entire field-of-view can be tiled out with coherently combined beams, which have the same sensitivity as imaging observations. Noise levels were based on expected values for the effective area and system temperature of operating and planned telescopes from the scientific literature and scaled assuming the integrated noise $\propto \sqrt{t_\mathrm{int}}$ (see Table~\ref{tab:telescopes} for the specific parameters used to simulate each observatory). 

The electron distribution in the IGM is not well known, but because the concentration of electrons is thought to be lower in the IGM, we anticipate that the scatter-broadening of pulses by the IGM may be lower than that of our Galaxy. Any contribution to the DM or scatter-broadening from a host galaxy will also be reduced because it will be redshifted from higher frequencies.
We therefore simulated two scenarios:
\begin{enumerate} 
\item \emph{High Scattering} -- The IGM causes temporal scatter broadening which follows the \cite{bcc+04} relation (the same as the ISM).
\item \emph{No Scattering} -- There is no temporal scatter broadening.
\end{enumerate} 
The real rate will lie between these two extremes. Where not specified, the Galactic DM is assumed to be 150~pc~cm$^{-3}$, which is the mean DM of all directions in the sky \citep[based on the NE2001 model,][]{cl02}. 

\subsection{Rate of the Parkes FRBs}
We calculate $\rho_0$ by applying our simulations to the surveys which found the known FRBs. The surveys concerned are the Parkes Magellanic Clouds Pulsar Survey \citep[MCS,][]{mfl+06},  the Parkes Multi-beam Pulsar Survey \citep[PMPS,][]{mlc+01}, and the High Time Resolution Universe Survey \citep[HTRU,][]{kjs+10}. All surveys were undertaken using the Parkes Multi-beam receiver (see Table~\ref{tab:telescopes} for instrumental properties). The MCS observed for a total of 480 hours, the PMPS observed for a total of 1800 hours, and at the time of the publication of \cite{tsb+13}, 552 hours of the HTRU had been searched for FRBs. One FRB was found in the MCS and the PMPS, and four were found in 24\% of the HTRU. We used our simulations to determine the volume of sky covered in each survey, and then used that to determine $\rho_0$. 

We find that in both of the proposed scattering scenarios, the MCS was limited by the maximum value of DM the data were dedispersed to, the DM range used was $0-500$~pc~cm$^{-3}$ ($\sim75$~pc~cm$^{-3}$ of which was Galactic), the survey was sensitive to $D_\mathrm{max} \sim 1.9$~Gpc. The field-of-view used in our simulations is significantly lower than that of \cite{lbm+07}, which we feel was overestimated. The Half Power Beam Width of a single beam of the Parkes Multi-beam receiver is $\sim14'$ \citep{swb+95}, thus the integrated surface area of a Gaussian function approximating a single beam is $\sim0.086$~deg$^2$, and the field-of-view of the full, 13-beam receiver is $\sim1.1$~deg$^2$. This leads to a slightly higher estimate of the event rate, $\rho_0=98_{-78}^{+225}$~Gpc$^{-3}$~day$^{-1}$.  In the high-scattering simulations, the PMPS was limited by scatter-broadening, and was sensitive out to $D_\mathrm{max} \sim 2.5$~Gpc, giving a rate of $12_{-9}^{+27}$~Gpc$^{-3}$~day$^{-1}$. In the no-scattering simulations, the PMPS was limited by the DM range searched, $0-2200$~pc~cm$^{-3}$ ($\sim250$~pc~cm$^{-3}$ of which was Galactic), $D_\mathrm{max} \sim 5.5$~Gpc, yielding a significantly lower value for the event rate, $\rho_0 = 1.1_{-0.9}^{+2.5}$~Gpc$^{-3}$~day$^{-1}$. Note that the agreement between the rates from the two surveys is significantly reduced when the IGM does not scatter-broaden the pulses, although the rates are roughly compatible in both cases. The HTRU was also limited by scatter-broadening in the high-scattering simulations ($D_\mathrm{max} \sim 2.5$~Gpc,  $\rho_0 \sim 143_{-42}^{+114}$~Gpc$^{-3}$~day$^{-1}$), and by the DM range searched in the no-scattering simulations ($D_\mathrm{max} \sim 5.2$~Gpc,  $\rho_0 \sim 17_{-5}^{+13}$~Gpc$^{-3}$~day$^{-1}$). The rates derived from the HTRU are more compatible with the rates from the MCS than the PMPS. The reason for this may be because most of the observations from the PMPS were in directions closer to the Galactic plane, where local scatter-broadening and dispersion are strongest. Thus, the true event rate may be closer to the high-scattering simulations for the PMPS, and the no-scattering simulations in the MCS and HTRU observations.

To find the combined rate of all of the Parkes surveys, we multiplied the Poissonian distributions implied from the calculated rates together, and found $\rho_0 = 51_{-14}^{+31}$~Gpc$^{-3}$~day$^{-1}$ for the high-scattering simulations and $5.3_{-1.4}^{+3.1}$~Gpc$^{-3}$~day$^{-1}$ for the no-scattering simulations. These are the values we use in all subsequent simulations. The true event rate lies somewhere between these two values, and from the observational evidence of the 6 known bursts, probably closer to the no-scattering simulations. So \citep[as was discussed in][]{lbm+07} this is compatible with rates for short GRBs and neutron star inspirals, but significantly lower than the rate of core-collapse supernovae unless the bursts are beamed (see Table~\ref{tab:rates}). We note that the DM range searched is the limiting factor in the no-scattering simulations for all surveys, so we predict that reprocessing the data out to a higher DM will yield new FRB detections.

\begin{table}
\caption{Comparison of transient rates.}
\begin{center}
\begin{tabular}{lcc}
\hline
Object & Rate$^\dagger$ 				& Reference\\
             & (~Gpc$^{-3}$~day$^{-1}$) 		&  \\
\hline
FRBs (high-scattering)	& $51_{-14}^{+31}$	& This work\\
FRBs (no-scattering)		& $5.3_{-1.4}^{+3.1}$& This work\\
Short GRBs			& $\sim 0.3 - 3 $	& \cite{fbm+12} \\
NS mergers			& $\sim 0.3 - 30$	& \cite{aaa+10} \\
CC Supernovae		& $\sim 200 - 2000$& \cite{lcl+11} \\
\end{tabular}
\end{center}
$^\dagger$ The rates given here are `local' ($z<1$), but the true rates depend on redshift. They should be treated as order of magnitude estimates. 
\label{tab:rates}
\end{table}

\subsection{Other Observatories}
\begin{table*}
\caption{Comparison of the parameters used for simulations of current and planned telescopes. The values of field-of-view and $A_\mathrm{eff}/T_\mathrm{sys}$ given here are calculated for the centre of the observing band listed. For LOFAR, we use noise levels derived from current transient imaging surveys, which may improve in the future.}
\begin{center}
\begin{tabular}{lccccc}
\hline
Telescope &  $A_\mathrm{eff}/T_\mathrm{sys}$ & $\nu_\mathrm{low}$ & $\nu_\mathrm{high}$ &  FoV & Reference\\
		 &  (m$^{2}$/K) & (MHz)	& (MHz) & (deg$^2$) \\
\hline
SKA-low & 5000 & 50 & 350 & 27 & \cite{dtm13}\\
SKA$_1$-low & 1000 & 50 & 350 & 27 & \cite{dtm13}\\
SKA-mid & 10000 & 1000 & 2000 & 0.5 & \cite{dtm13}\\
SKA$_1$-mid & 1630 & 1000 & 2000 & 0.5 & \cite{dtm13}\\
LOFAR-HBA & 110 & 155 & 165 & 150 &\cite{sha+11}, \cite{vwg+13} \\
LOFAR-LBA & 0.5 & 30 & 80 & 100 & \cite{sha+11}, \cite{vwg+13} \\
MWA & 13.0 & 185 & 215 & 375 & \cite{tgb+13} \\
ASKAP & 81 & 700 & 1000 & 30 & \cite{jfg09}\\
MeerKAT & 220 & 580 & 1750 & 1.0 & \cite{bbj10} \\
Parkes Multi-beam & 92 & 1230 & 1518 & 1.1 & \cite{mlc+01}\\
Molonglo & 277 & 790 &  890 &  12 & \cite{gmc+12} \\
UTR-2 & 0.5 & 10 &  20 &  40 & \cite{abzk01}\\
LWA & 30 & 50 & 70 & 20 & \cite{ecc+09} \\
\end{tabular}
\end{center}
\label{tab:telescopes}
\end{table*}%

\begin{figure*}
\begin{center}
\includegraphics[height=\linewidth, angle=270, trim=3.0cm 0.5cm 3.0cm 0.5cm, clip=True]{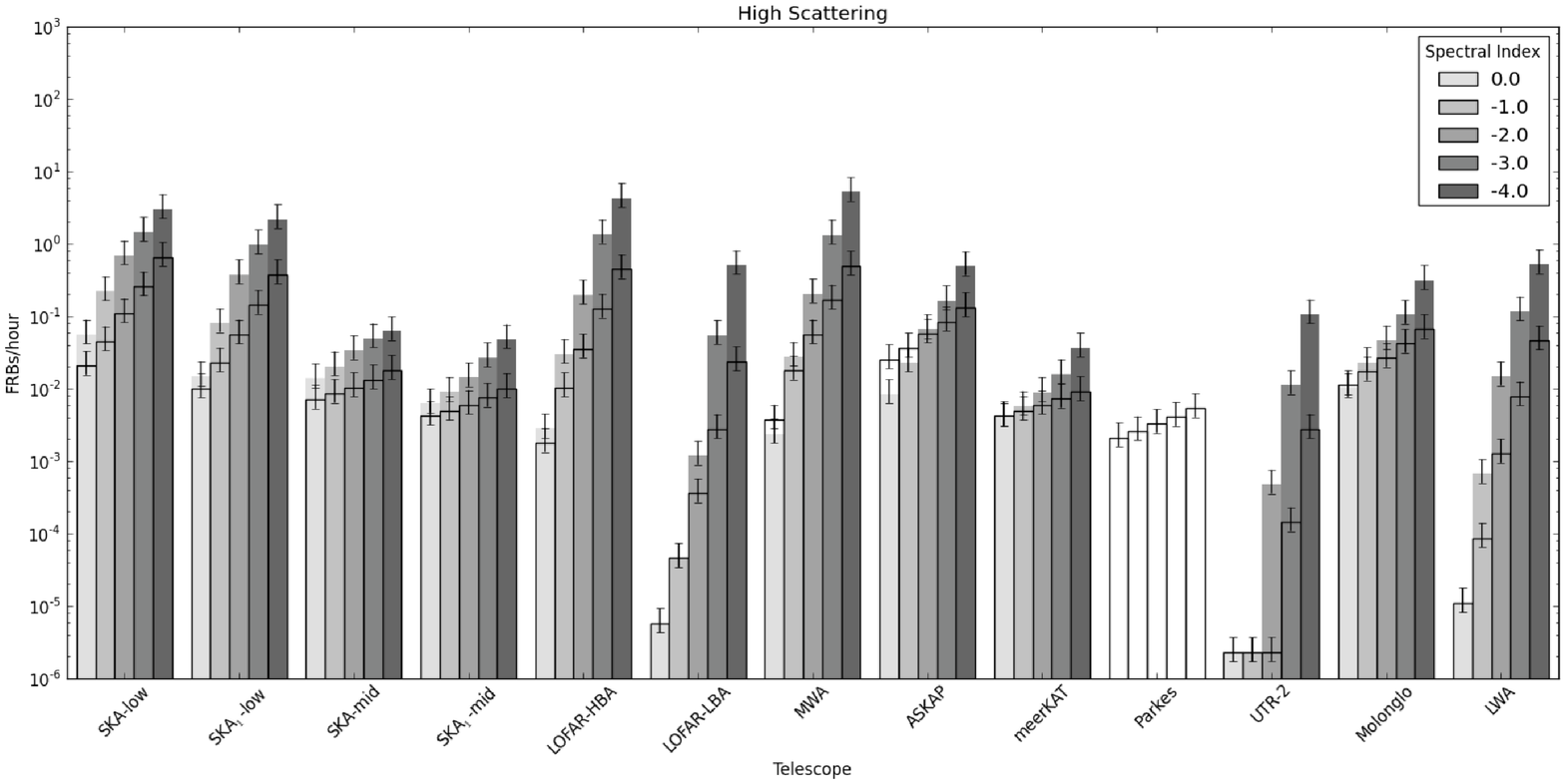}
\caption{Expected number of FRBs per hour for various observatories in the high-scattering simulations. The coloured bars show the number of FRBs detectable in imaging surveys, assuming different spectral indices of: 0.0 (white), -1.0, -2.0, -3.0 and -4.0 (darkest grey). The number of FRBs detectable in beamformed surveys are indicated by the bars with a solid black outline. The DM range used was $0-6000$~pc~cm$^{-3}$.}
\label{fig:numbers}
\includegraphics[height=\linewidth, angle=270, trim=3.0cm 0.5cm 3.0cm 0.5cm, clip=True]{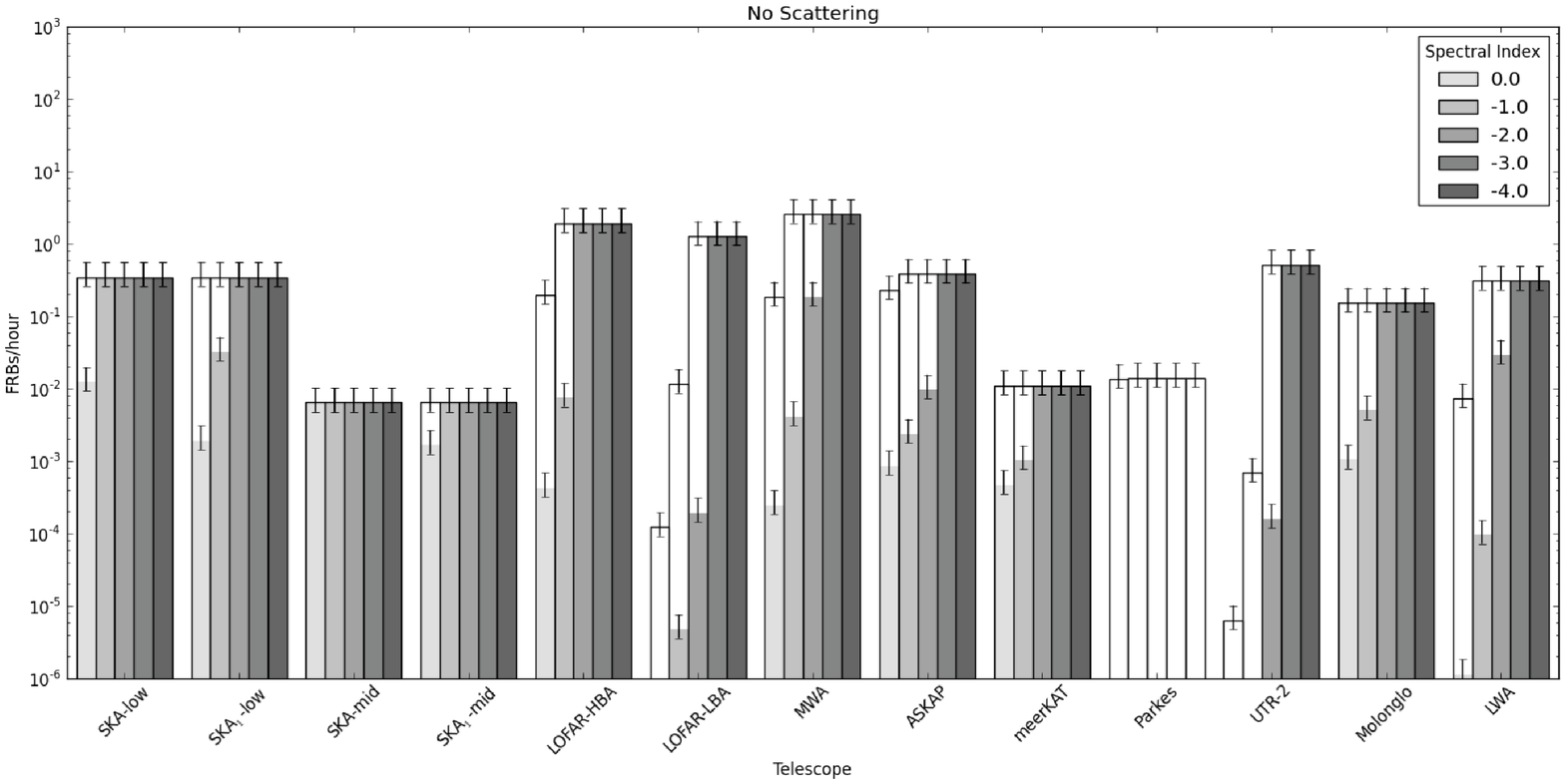}
\caption{As Figure~\ref{fig:numbers}, but for the no-scattering simulations.}
\label{fig:numbers_noscatt}
\end{center}
\end{figure*}

Using the rates derived from the Parkes surveys, we applied our simulations to other existing and planned observatories to determine which will be most suitable for finding FRBs. The observatories considered, and the observing parameters used in the simulations are summarised in Table~\ref{tab:telescopes}. We note that even if the occurence rate changes significantly as more FRBs are discovered, the relative performance of the telescopes shown in these plots will remain accurate, and the detectable numbers will simply be scaled by a constant factor. Figures~\ref{fig:numbers} and \ref{fig:numbers_noscatt} show the results of the simulations for the high-scattering and no-scattering simulations respectively at each observatory. The coloured bars show the number of FRBs detectable in imaging surveys, assuming different spectral indices of: 0.0 (white), -1.0, -2.0, -3.0 and -4.0 (darkest grey). The number of FRBs detectable in beamformed surveys are indicated by the bars with a solid black outline. These results are also tabulated in Appendix~\ref{sec:tables}.

It is clear from the figures that in the low-scattering simulations, beamformed observations are typically more efficient at finding FRBs than imaging observations, however, in telescopes which are very sensitive (eg. SKA-mid), imaging surveys are competitive. In the high-scattering simulations, imaging surveys are often much more effective. We note that, as the amount of scatter-broadening is not known a priori, both imaging \emph{and} beamformed observations are needed to maximise chances of detection. One can also see that the field-of-view of the telescope makes a big difference. This is most obvious when comparing ASKAP and the Parkes telescope in the no-scattering beamformed simulations. Whilst both telescopes observe at similar frequencies,  ASKAP is slightly less sensitive, but should find many more FRBs because of its much larger field-of-view (although this relies on the assumption that the enitre field-of-view can be tiled out with coherently-added beams in beamformed mode). MeerKAT also observes at the same frequency, is far more sensitive, and has a much wider bandwidth, but because the field-of-view is only one square degree, it will find FRBs at a slightly lower rate than Parkes, and at a significantly lower rate than ASKAP.

\section{Search Strategies}
\subsection{Matched Filtering}
\label{sec:opt}
Matched filtering is used to maximise the SNR of a given dataset. This is done by correlating noisy data with a noiseless template approximating the shape of the underlying signal being searched for. Since the shape, and width of the pulse are not known a priori this is most simply done by iteratively summing adjacent samples. This technique is commonly used in single pulse searches of beamformed data \citep{cm03}, and should also be applied when searching for transients in images. Images should be searched on a range of timescales (i.e. varying integration times) in order to tune observations to pulses of different widths.  

For scattered bursts, the shape of the signal at a given frequency is an exponentially-decaying pulse, $e^{-t/\tau_\mathrm{s}}$. The integrated signal of the pulse increases as $\tau_\mathrm{s}(1-e^{-t_\mathrm{int}/\tau_\mathrm{s}})$, whilst the integrated noise level increases as $\sqrt{t_\mathrm{int}}$. Therefore, the SNR as a function of integration time, $t_\mathrm{int}$, scales as:
\begin{equation}
\mathrm{SNR} \propto \frac{\tau_\mathrm{s}(1-e^{-t_\mathrm{int}/\tau_\mathrm{s}})}{\sqrt{t_\mathrm{int}}}~,
\end{equation}
the peak of this function occurs when $t_\mathrm{int} \approx \tau_\mathrm{s}$. This is complicated by the frequency evolution of the pulse, but from our simulations, we find that generally, as adding noise only reduces sensitivity by $1/\sqrt{t_\mathrm{int}}$, the optimum integration time is approximately equal to the scatter-broadening time at the bottom of the band. Using a range of integration times makes the data sensitive to pulses of different durations. This part of the search process is particularly important for low frequency surveys, where scatter broadening means that pulse-widths can range from a few milliseconds to several hours. 

\subsection{Line-of-Sight}
\begin{figure}
\begin{center}
\includegraphics[height=\linewidth, angle=270, trim=0.5cm 2.5cm 0.5cm 2.25cm, clip=True]{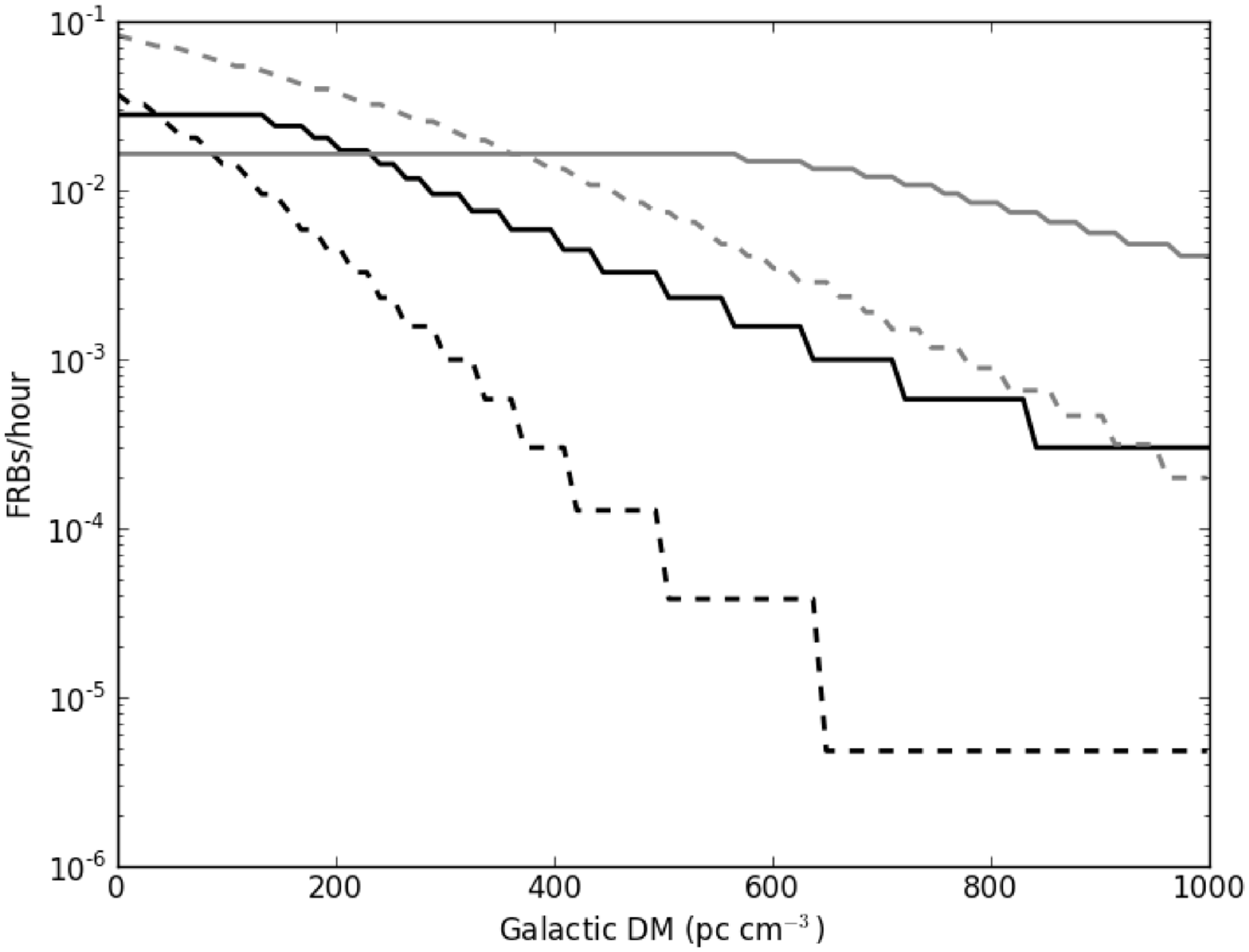}
\caption{The number of detectable FRBs per hour as seen by the LOFAR HBAs (black lines) and ASKAP (grey lines). The solid lines correspond to imaging observations, and the dashed lines correspond to beamformed observations. The spectral index used for these simulations is $\alpha=-2.0$, but changing the spectral index only affects the scale of the figure, and does not impact shape of the curves significantly. The observing parameters of both observatories are given in Table~\ref{tab:telescopes}.}
\label{fig:dms}
\end{center}
\end{figure}

The specific lines-of-sight used will have a significant impact on the sensitivity of a transient survey. Scattering significantly reduces the peak flux of a pulse, so maximal sensitivity to extragalactic sources is achieved when observing away from the Galactic plane ($|b|\gtrsim 10\degr$), where the electron density along the line-of-sight, and therefore the scatter-broadening time, is lowest. This will impact all radio transient surveys, but is particularly important at low frequencies. Figure~\ref{fig:dms} shows the number of FRBs expected to be observed per hour using the LOFAR HBAs (black lines) and ASKAP (grey lines) in our high-scattering simulations as a function of the Galactic DM along the line-of-sight (see Table~\ref{tab:telescopes} for the specifications of the telescopes used in the simulation). The contribution from the host galaxy is assumed to be small, but we do consider the contribution from the IGM. The solid lines show the rates for imaging observations, and the dashed lines show the rates for beamformed observations. For LOFAR beamformed observations, choosing a `clear' line-of-sight can improve the chances of detecting an FRB by a factor of $\sim10^4$. For ASKAP (a higher frequency instrument), the effect is smaller, but still makes a significant difference. The effect is less pronounced in imaging observations, but still important. It should be noted that the simulations used to produce Figure~\ref{fig:dms} do not include the lever-arm effect, and because this will reduce scatter-broadening, the trend may be slightly exaggerated. However, even in the no-scattering simulations, choosing a line-of-sight with a low Galactic DM still increases the volume of extragalactic space probed.

\subsection{Dedispersion}
\begin{figure}
\begin{center}
\includegraphics[height=\linewidth, angle=90, trim=0.5cm 2.5cm 0.5cm 2.25cm, clip=True]{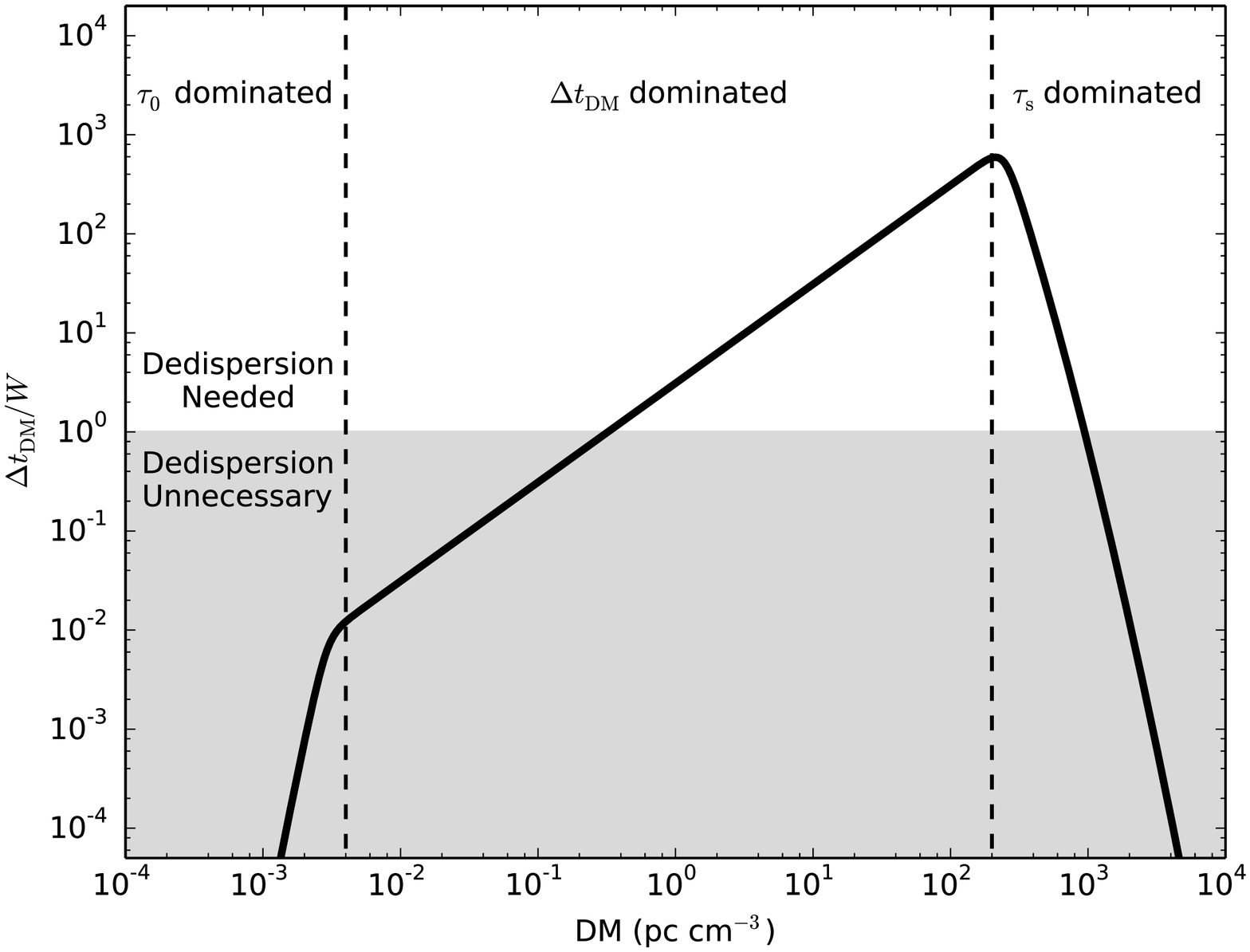}
\caption{$\Delta t_\mathrm{DM}/W$ plotted as a function of DM for a pulse detected at a central frequency of 1500~MHz with 1000~MHz bandwidth in the high-scattering regime. Dedispersion is necessary when $\Delta t_\mathrm{DM}/W > 1$ (the white area of the plot), but becomes unnecessary when $\Delta t_\mathrm{DM}/W < 1$ (the grey area of the plot).}
\label{fig:dedisp}
\end{center}
\end{figure}

Dedispersion is only beneficial for detecting FRBs when the dispersive delay across the band ($\Delta t_\mathrm{DM}$) is larger than the observed width of the pulse ($W$). Figure~\ref{fig:dedisp} shows $\Delta t_\mathrm{DM}/W$ as a function of DM for a pulse detected at a central frequency of 1500~MHz with 1000~MHz bandwidth in the high-scattering simulations. At low DMs, $W$ is dominated by the intrinsic width of the pulse and it is not necessary to dedisperse the data because $\Delta t_\mathrm{DM}$ is small. $\Delta t_\mathrm{DM}$ increases with DM until it exceeds the intrinsic pulse width, at which point dedispersion becomes necessary. This is the regime which FRBs (and pulsars) are normally detected in beamformed surveys. However, above a certain DM, the scatter-broadening of the pulse, which increases more rapidly with DM than the dispersive delay, means that $W$ eventually exceeds $\Delta t_\mathrm{DM}$ again. Thus, for highly-scattered sources, it is not beneficial to dedisperse the data. It is these highly-scattered objects which would be the targets of imaging surveys for FRBs. At low frequencies, or with narrower bandwidths, the critical DM at which scatter broadening exceeds  $\Delta t_\mathrm{DM}$ is shifted to lower DMs. For example, using the maximum bandwidth available with the LOFAR high band antennas (96~MHz), with a central frequency of 150~MHz, the SNR of observations is not improved by dedispersion for DMs above $\sim$400~pc~cm$^{-3}$. For the LOFAR low band antennas (80~MHz bandwidth, centred on 60~MHz), dedispersion does not enhance our sensitivity for DMs above $\sim$250~pc~cm$^{-3}$. The bursts we are interested in occur outside our Galaxy, which means they typically have DMs $>150$~pc~cm$^{-3}$ from the Galactic dispersion along the line-of-sight, and the dispersion from within their host galaxy. So, although LOFAR imaging observations will not be dedispersed, that should not impact their chances of detecting such bursts significantly. 

\subsection{Imaging vs. Beamformed Observations}
Imaging surveys are competitive with beamformed surveys in finding FRBs when the scattering of the pulse is significant. This is because images are inherently more robust against RFI (affected stations can be removed and spurious sources can be rejected when they do not appear point-like, although with better RFI-removal algorithms the difference between imaging and beamformed data could be reduced), and more stable on long timescales. In addition, because the more distant elements of an array are easier to include in imaging observations, finding FRBs in images could offer a much better localisation of the source, and could help to associate it with a host galaxy. Unfortunately, producing images which have integration times shorter than a few minutes is often difficult because short integrations have reduced UV-coverage, which can lead to difficulties calibrating and cleaning the data correctly. Exceptions to this include arrays with good instantaneous UV-coverage, and situations where the burst dominates the flux in the field-of-view. Also, there is a practical limit on the shortest images, which is set by the shortest possible correlator time, so images are less sensitive to short bursts, and may be unable to resolve scatter broadening or dispersion \citep[although this limit has been improved radically, see][]{ljb+11}.  

This means that imaging observations are not as efficient as beamformed observations for short-duration pulses. In addition, dedispersion significantly increases the SNR for short pulses, and becomes vital {when scatter-broadening is small}. Because of the amount of extra processing required, this is impractical on imaging datasets but is easily manageable on beamformed data, which tend to have much lower data-rates. When scatter-broadening is important, however, imaging observations are much more robust. It is easy to introduce long-timescale fluctuations into the baselines of beamformed data, which reduce the sensitivity to very scattered bursts. For example, with LOFAR, bright sources moving in and out of the sidelobes of the beam can cause the baseline to fluctuate on timescales of a few minutes. Where possible, fast transient surveys should perform observations with both imaging \emph{and} beamformed data as both modes probe different areas of pulse-width parameter-space.

\section{Conclusions}
The next generation of radio telescopes should find a large population of FRBs. In the most optimistic scenario, the low-frequency component of SKA$_1$ could find up to 1 FRB per hour. However, FRB searches must be optimised, in order to increase the probability of detection. The instantaneous field-of-view should be as large as possible, FRBs are bright and do not require much sensitivity to detect, but they are quite rare, so probing a large volume of space is  beneficial. This volume can in some circumstances be limited by the range of DMs searched, so processing to very high DMs (e.g. up to at least 6000) will maximize new discoveries. The line-of-sight through the Galaxy should be chosen carefully to minimise any local contribution to scattering and dispersion. Finally, matched filtering should be employed in both imaging and beamformed surveys of transients -- the sampling time should be matched to the duration of the bursts to maximise detection. In the case of typical imaging surveys, this means that the data should be imaged on a range of timescales from seconds to hours. Imaging surveys for FRBs in some circumstances can be competitive with beamformed observations, and it should be possible to detect FRBs in imaging surveys for `slow' transients. Any such detections would be very useful because of the potential to identify the host of the event. We also note that imaging surveys probe different pulse widths than those of beamformed surveys, and could be effective at detecting the potential population of highy-scattered FRBs. As the extent of the scatter-broadening is not known a priori, the most effective use of telescope time would involve simultaneous beamformed and imaging surveys.

\section*{Acknowledgements}
This project was funded by European Research Council Advanced Grant 267697 ``4 Pi Sky: Extreme Astrophysics with Revolutionary Radio Telescopes''.

\bibliography{refs}
\bibliographystyle{mn2e}

\appendix
\section{Tables of Derived Rates}
\label{sec:tables}
Here we tabulate the results of our simulations. Table~\ref{tab:rates_bf} shows the number of FRBs expected per hour in our simulations of beamformed observations, and Table~\ref{tab:rates_im} shows the number of FRBs expected per hour from our simulations of imaging observations. In each table, the rates are given for both the no-scattering and the high-scattering simulations. 

\begin{table*}
\begin{minipage}{17.5cm}
\caption{Derived rates for the numbers of FRBs in no-scattering/high-scattering simulations of  beamformed observations.}
\centering
\begin{tabular}{lccccc}
\hline
Instrument	& \multicolumn{5}{c}{FRBs/hour derived from no-scattering / high-scattering simulations} \\
	    		& $\alpha=0$ & $\alpha=-1$ & $\alpha=-2$ & $\alpha=-3$ & $\alpha=-4$ \\\hline
SKA-low &  0.35 /  0.021 &  0.35 /  0.045 &  0.35 /  0.11 &  0.35 /  0.26 &  0.35 /  0.66 \\ 
SKA$_1$-low &  0.35 /  0.01 &  0.35 /  0.023 &  0.35 /  0.056 &  0.35 /  0.14 &  0.35 /  0.38 \\ 
SKA-mid &  0.0064 /  0.007 &  0.0064 /  0.0085 &  0.0064 /  0.01 &  0.0064 /  0.013 &  0.0064 /  0.018 \\ 
SKA$_1$-mid &  0.0064 /  0.0042 &  0.0064 /  0.0049 &  0.0064 /  0.0059 &  0.0064 /  0.0075 &  0.0064 /  0.01 \\
LOFAR-HBA &  0.2 /  0.0018 &  1.9 /  0.01 &  1.9 /  0.036 &  1.9 /  0.13 &  1.9 /  0.45 \\ 
LOFAR-LBA &  0.00012 /  5.8e-06 &  0.012 /  4.6e-05 &  1.3 /  0.00036 &  1.3 /  0.0028 &  1.3 /  0.024 \\ 
MWA &  0.19 /  0.0037 &  2.6 /  0.018 &  2.6 /  0.056 &  2.6 /  0.17 &  2.6 /  0.5 \\ 
ASKAP &  0.23 /  0.025 &  0.38 /  0.037 &  0.38 /  0.058 &  0.38 /  0.084 &  0.38 /  0.13 \\
meerKAT &  0.011 /  0.0042 &  0.011 /  0.0049 &  0.011 /  0.006 &  0.011 /  0.0073 &  0.011 /  0.0092 \\
Parkes &  0.014 /  0.0021 &  0.014 /  0.0026 &  0.014 /  0.0033 &  0.014 /  0.004 &  0.014 /  0.0054 \\
UTR-2 &  6.3e-06 /  2.3e-06 &  0.00069 /  2.3e-06 &  0.51 /  2.3e-06 &  0.51 /  0.00014 &  0.51 /  0.0028 \\
Molonglo &  0.15 /  0.011 &  0.15 /  0.017 &  0.15 /  0.026 &  0.15 /  0.042 &  0.15 /  0.066 \\ 
LWA &  0.0074 /  1.1e-05 &  0.31 /  8.6e-05 &  0.31 /  0.0013 &  0.31 /  0.0078 &  0.31 /  0.046 \\ \hline
\end{tabular}
\end{minipage}
\label{tab:rates_bf}
\end{table*}%

\begin{table*}
\begin{minipage}{17.5cm}
\caption{Derived rates for the numbers of FRBs in no-scattering/high-scattering simulations of  imaging observations.}
\centering
\begin{tabular}{lccccc}
\hline
Instrument	& \multicolumn{5}{c}{FRBs/hour derived from no-scattering / high-scattering simulations} \\
	    		& $\alpha=0$ & $\alpha=-1$ & $\alpha=-2$ & $\alpha=-3$ & $\alpha=-4$ \\
\hline
SKA-low &  0.012 /  0.056 &  0.35 /  0.22 &  0.35 /  0.69 &  0.35 /  1.5 &  0.35 /  3.0 \\ 
SKA$_1$-low &  0.0019 /  0.015 &  0.032 /  0.08 &  0.35 /  0.38 &  0.35 /  0.98 &  0.35 /  2.2 \\
SKA-mid &  0.0064 /  0.014 &  0.0064 /  0.02 &  0.0064 /  0.034 &  0.0064 /  0.05 &  0.0064 /  0.062 \\ 
SKA$_1$-mid &  0.0017 /  0.0063 &  0.0064 /  0.009 &  0.0064 /  0.014 &  0.0064 /  0.027 &  0.0064 /  0.048 \\
LOFAR-HBA &  0.00043 /  0.0028 &  0.0075 /  0.03 &  1.9 /  0.2 &  1.9 /  1.4 &  1.9 /  4.3 \\
LOFAR-LBA &  6e-07 /  5.8e-06 &  4.7e-06 /  4.6e-05 &  0.00019 /  0.0012 &  1.3 /  0.055 &  1.3 /  0.51 \\
MWA &  0.00025 /  0.0024 &  0.0041 /  0.028 &  0.19 /  0.2 &  2.6 /  1.3 &  2.6 /  5.2 \\
ASKAP &  0.00086 /  0.0084 &  0.0024 /  0.023 &  0.0097 /  0.066 &  0.38 /  0.17 &  0.38 /  0.49 \\
meerKAT &  0.00047 /  0.004 &  0.001 /  0.0057 &  0.011 /  0.009 &  0.011 /  0.016 &  0.011 /  0.037 \\
Parkes &  - &  - &  - &  - &  - \\
UTR-2 &  2.4e-07 /  2.3e-06 &  2.4e-07 /  2.3e-06 &  0.00016 /  0.00048 &  0.51 /  0.011 &  0.51 /  0.11 \\
Molonglo &  0.001 /  0.01 &  0.005 /  0.023 &  0.15 /  0.046 &  0.15 /  0.1 &  0.15 /  0.32 \\
LWA &  1.1e-06 /  1.1e-05 &  9.6e-05 /  0.00066 &  0.029 /  0.015 &  0.31 /  0.12 &  0.31 /  0.52 \\
\hline
\end{tabular}
\end{minipage}
\label{tab:rates_im}
\end{table*}%

\label{lastpage}
\end{document}